\documentstyle[epsf,pennames]{prhep97}

\makeatletter
\let\chapter\hid@chapter
\makeatother
\newcommand{\alphas}  {\mbox{$\alpha_{\mathrm{s}}$}}
\newcommand{\pbinv}  {\mbox{~${\mathrm{pb}^{-1}}$}}
\newcommand{\MW}  {\mbox{$M_{\mathrm{W}}$}}
\newcommand{\MH}  {\mbox{$m_{\mathrm{H}}$}}
 \def\PZ{\relax\ifmmode{\rm Z}%
         \else${\rm W }$\fi}%
\newcommand{\MZ}  {\mbox{$M_{\mathrm{Z}}$}}
\newcommand{\Mt}  {\mbox{$m_{\mathrm{t}}$}}
\newcommand{\Vcs}  {\mbox{$V_{\mathrm{cs}}$}}
\newcommand{\stlef}  {\mbox{$\sin^2\theta^{\mathrm{lept}}_{\mathrm{eff}}$}}
\begin{document}
%\pagenumbering{empty}

% The following definitions need to be customised;

% Will appear on page headings
\authorrunning{D.R.~Ward}
\titlerunning{{\talknumber}: Tests of the Standard Model $\ldots$}
 
% Now the full name of author and talk

% For plenary talks, the talk number is that of the session
\def\talknumber{15} 

\title{{\talknumber}: Tests of the Standard Model: W mass and WWZ Couplings }
\author{David\,Ward
%\inst{1}
(drw1@cam.ac.uk)}
\institute{ Cavendish Laboratory, University of Cambridge, U.K.}

\maketitle

\vspace{-6cm}
\begin{flushright}
\hspace{2cm} \\  14 November 1997 \\
\end{flushright}
\vspace{4.1cm}
{\footnotesize \it Plenary talk at the International Europhysics
       Conference on High Energy Physics, Jerusalem, 19--26 August 1997}
\begin{abstract}
Recent tests of the electroweak Standard Model are reviewed, covering the 
precise measurements of Z decays at LEP~I and SLC and
measurements of fermion pair production at higher energies at LEP~II.
Special emphasis is given to new results on W physics from LEP and FNAL.
\end{abstract}
\section{Precision measurements of the Z boson}
\subsection{Lineshape and leptonic forward-backward asymmetries}
The accurate measurement of the Z mass is essential for precise 
tests of the Standard Model.  This measurement is dominated by the 
energy scans performed at LEP in 1993 and 1995, in which approximately 
40\pbinv\ of data were recorded at energies $\pm1.8$~GeV away from 
the Z peak.  Important progress has been made in the last year in 
understanding the LEP beam energy for these scans, and final results
became available shortly before the conference.
The estimated errors on the centre-of-mass energy (in MeV) are: 
\begin{center}
\begin{tabular}{c@{\hspace{2em}}c@{\hspace{2em}}c@{\hspace{2em}}c}
\hline
         &  peak$-$2  &  peak  &  peak+2 \\
\hline
1993     &    3.5     &  6.7   &   3.0   \\
1994     &            &  3.7   &         \\
1995     &    1.8     &  5.4   &   1.7   \\
\hline 
\end{tabular}
\end{center}

The path is now clear for the LEP experiments to complete their analyses 
of the data.  The principal changes to the results 
compared to those presented in 1996~\cite{blondel,lepew96} 
are (see~\cite{quast} for more details):
\begin{itemize}
\item[$\bullet$]
 The new beam energy values have already been incorporated by ALEPH and
DELPHI; 
OPAL and L3 have chosen not to update their results at this stage, but 
an appropriate correction to the Z mass and width values has been applied.
\item[$\bullet$] 
Some data from the 1995 run have been added by OPAL and L3, 
and there have been significant changes to the $\tau^+\tau^-$ measurements
from ALEPH.
\item[$\bullet$]
L3 have included the 1995 data in their $\tau$ polarization 
measurements~\cite{pr481}.
\item[$\bullet$] 
A preliminary measurement of $A_{\mathrm{LR}}$ from 
the SLD 1996 data is available.
\end{itemize}
Overall the changes in the combined results since last year are quite small.  
The basic combined LEP measurements are~\cite{quast}:
\begin{center}
\begin{tabular}{c@{\hspace{3em}}c}
Without lepton universality & Assuming lepton universality  \\
\begin{tabular}{c@{\hspace{2em}}c}
\hline
$\MZ$ /GeV              & $91.1867\pm0.0020$ \\
$\Gamma_{\mathrm{Z}}$ /GeV         & $2.4948\pm0.0025$ \\
$\sigma_{\mathrm{h}}^0$ /nb & $41.486\pm0.053 $ \\
$R_{\Pe}$                  & $20.757\pm0.056 $ \\
$R_{\Pgm}$                  & $20.783\pm0.037 $ \\
$R_{\Pgt}$                  & $20.823\pm0.050 $ \\
$A_{\mathrm{FB}}^{0,\Pe}$  & $0.0160\pm0.0024$ \\
$A_{\mathrm{FB}}^{0,\Pgm}$  & $0.0163\pm0.0014$ \\
$A_{\mathrm{FB}}^{0,\Pgt}$  & $0.0192\pm0.0018$ \\
\hline
\end{tabular}
& 
\begin{tabular}{c@{\hspace{2em}}c}
\hline
$\MZ$ /GeV              & $91.1867\pm0.0020$ \\
$\Gamma_{\mathrm{Z}}$ /GeV         & $2.4948\pm0.0025$ \\
$\sigma_{\mathrm{h}}^0$ /nb & $41.486\pm0.053 $ \\
 & \\
$R_{\ell}$                  & $20.775\pm0.027 $ \\
 & \\
 & \\
$A_{\mathrm{FB}}^{0,\ell}$  & $0.0171\pm0.0010$ \\
 & \\
\hline
\end{tabular}
\\ $\chi^2$/dof=21/27 & $\chi^2$/dof=23/31 \\
\end{tabular}
\end{center}
The data from the four experiments are in excellent agreement, 
as shown by the values of $\chi^2$/dof.

From these measurements, it is possible to infer a value
for the invisible width of the Z:
\[ \Gamma_{\mathrm{inv}}=500.1\pm1.8\;\mathrm{MeV} \]
which can be converted into a measurement of the number of light neutrino
species assuming they have Standard Model couplings:
\[  N_{\nu} = 2.993 \pm 0.011 \]
A limit on the possible additional invisible width arising from new physics
can also be inferred: $\Delta\Gamma_{\mathrm{inv}}<2.8$~MeV at 95\% c.l. 

\begin{figure}
\centering\parbox[t]{.48\textwidth}{\centering\leavevmode
\epsfysize=6cm
\centerline{\epsffile{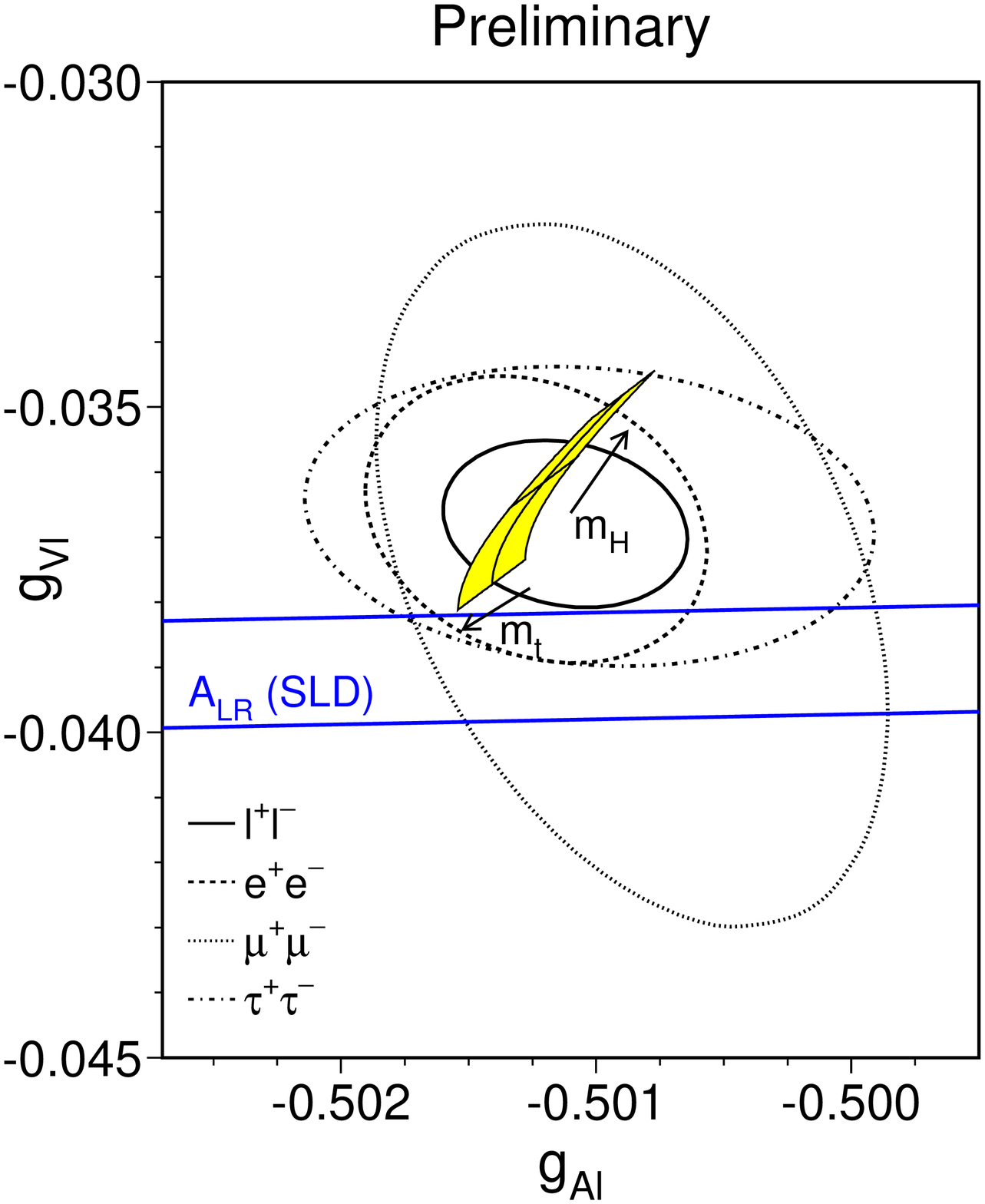}}
\caption{68\% probability contours of vector and axial vector 
couplings of the Z to leptons.}
\label{fig_leptuniv}
}
\hfill
\centering\parbox[t]{.48\textwidth}{\centering\leavevmode
\epsfysize=6cm
\centerline{\epsffile{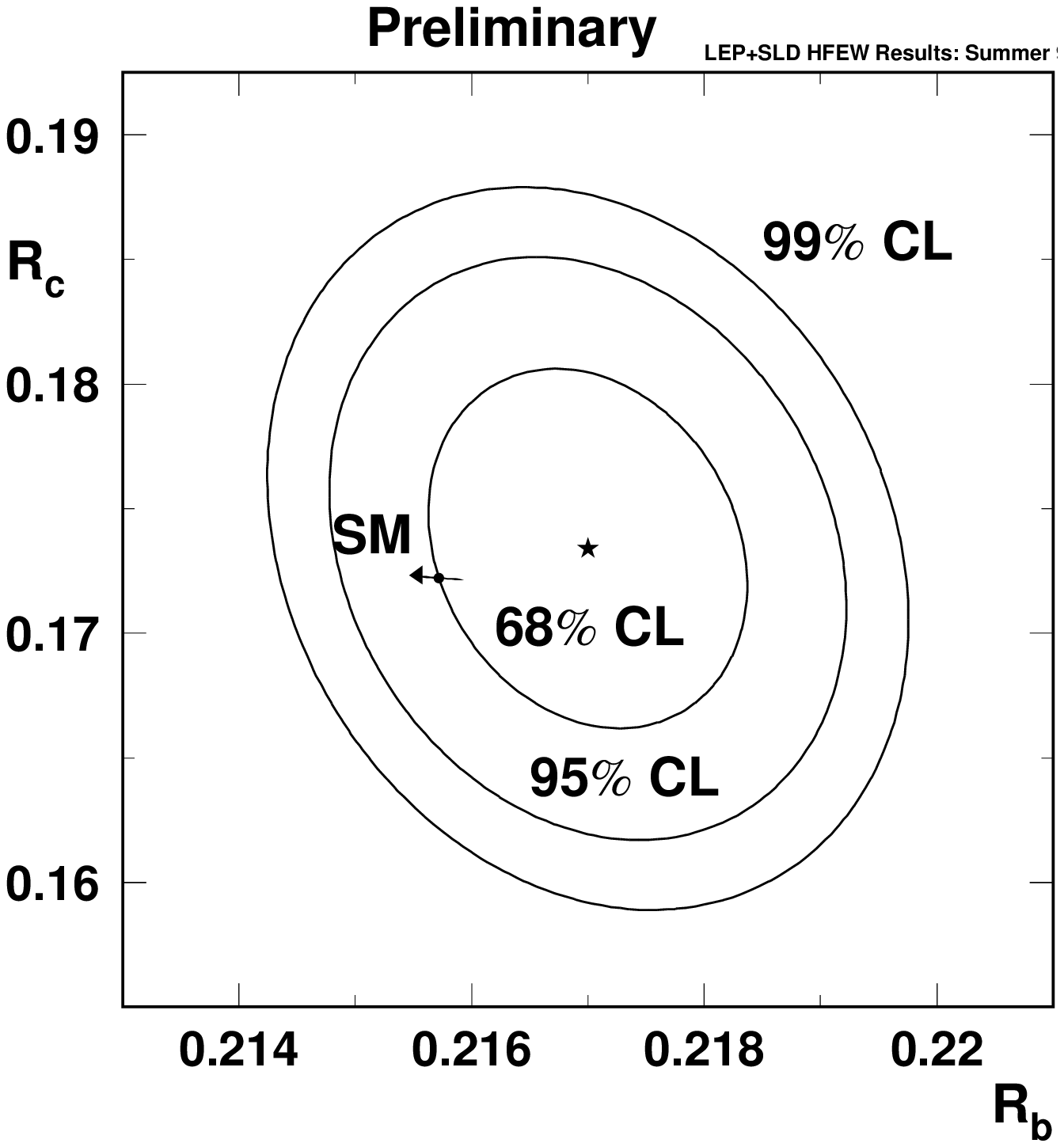}}
\caption{Measurements of $R_{\mathrm{b}}$ and $R_{\mathrm{c}}$, compared with
the Standard Model.}
\label{fig_rbrc}
}
\end{figure}

The results from the Z leptonic lineshape and asymmetries can be combined
with measurements of the $\tau$ polarization and its asymmetry 
(which are sensitive to the electron and $\tau$ couplings separately)
to perform the best tests of lepton universality.
In fig.~\ref{fig_leptuniv} we show the vector and axial couplings for each
lepton species, as inferred from the LEP data.  
The measurements are clearly consistent with universality of the   
Z leptonic couplings, with a precision of $\sim$0.2\% for $g_{\mathrm{A}}$ and
5-10\% for $g_{\mathrm{V}}$, and also with the Standard Model expectation, 
indicated by the shaded area with uncertainties arising from 
varying the Higgs mass between 60 and 1000~GeV and the t-quark mass
in the range $175.6\pm5.5$~GeV.  The measurement of $A_{\mathrm{LR}}$ from SLD
is also shown.  Under the assumption of lepton universality, the 
following values for the couplings are obtained:
\begin{center}
\begin{tabular}{c@{\hspace{2em}}c@{\hspace{2em}}c}
\hline
           & LEP & LEP+SLD \\       
\hline
$g_{\mathrm{V}\ell}$ & $-0.03681\pm0.00085$ & $-0.03793\pm0.00058$ \\
$g_{\mathrm{A}\ell}$ & $-0.50112\pm0.00032$ & $-0.50103\pm0.00031$ \\
\hline
\end{tabular}                                    
\end{center}

\subsection{Heavy flavour electroweak measurements}
The main changes in the past year, reviewed in~\cite{mariotti,etzion}, are:
\begin{itemize}
\item[$\bullet$]
Measurements of $R_{\mathrm{b}}=\Gamma_{\mathrm{b}\overline{\mathrm{b}}}/
\Gamma_{\mathrm{had}}$ from ALEPH~\cite{A_Rb},  
OPAL~\cite{O_Rb} and SLD have been finalised for
publication in the last year, and new preliminary measurements from
DELPHI~\cite{pr419}, L3~\cite{pr489} and SLD~\cite{pr118} 
have led to significantly improved precision.
\item[$\bullet$]
New determinations of $R_{\mathrm{c}}$ from OPAL~\cite{O_Rc}, 
ALEPH~\cite{pr623} and SLD~\cite{pr120} (the latter 
exploiting a double vertex tag), have led to continued improvement in
the precision of this measurement.                            
\item[$\bullet$]
There are new measurements of the forward backward asymmetry 
$A_{\mathrm{FB}}^{\mathrm{b}}$ from L3~\cite{pr490} and OPAL~\cite{O_AFBb}, 
but an ALEPH result has been
withdrawn, so the overall precision of the measurement is unchanged.
\item[$\bullet$]
The SLD measurements of the polarized asymmetries~\cite{pr122}
 have led to a much improved determination of $\mathcal{A}_{\mathrm{c}}$.
\end{itemize}
The combined LEP/SLD heavy flavour measurements may be summarized as 
follows:
\begin{center}
\begin{tabular}{c@{\hspace{2em}}c}
\hline
$R_{\mathrm{b}}$                & $0.2170 \pm0.0009$ \\
$R_{\mathrm{c}}$                & $0.1734 \pm0.0048$ \\
$A_{\mathrm{FB}}^{0,\mathrm{b}}$ & $0.0984 \pm0.0024$ \\
$A_{\mathrm{FB}}^{0,\mathrm{c}}$ & $0.0741 \pm0.0048$ \\
$\mathcal{A}_{\mathrm{b}}$      & $0.900  \pm0.050 $ \\
$\mathcal{A}_{\mathrm{c}}$      & $0.650  \pm0.058 $ \\
\hline
\end{tabular}
\end{center}
In fig.~\ref{fig_rbrc} we compare the measured values of 
$R_{\mathrm{b}}$ and $R_{\mathrm{c}}$ with the Standard Model
expectations.  The apparent disagreement which excited much interest
in previous years has evaporated.
 
\begin{figure}
\centering\parbox[t]{.48\textwidth}{\centering\leavevmode
\epsfysize=6cm
\centerline{\epsffile{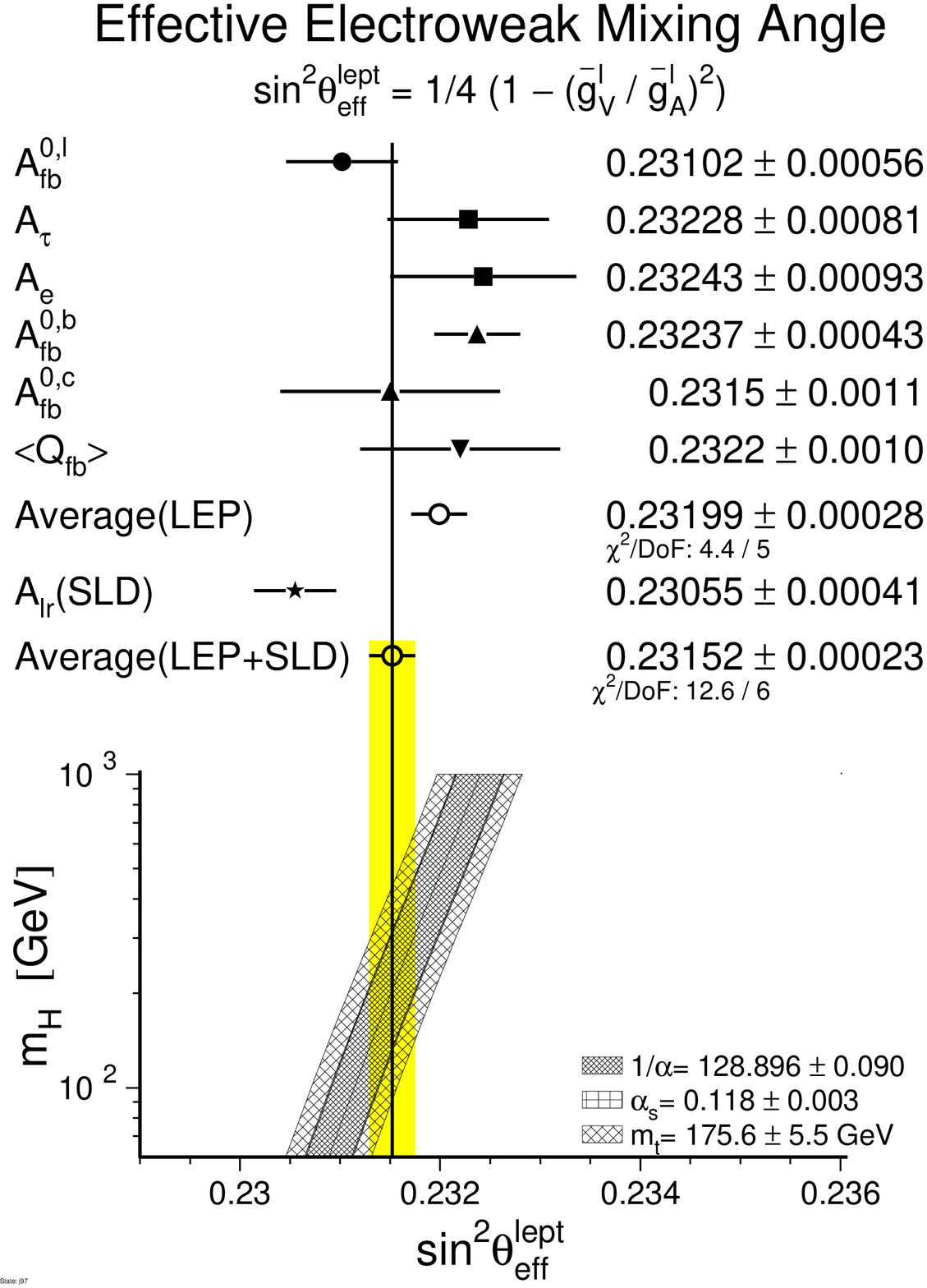}}
\caption{Determinations of \stlef\ from asymmetry and $\tau$ polarization
measurements.  The Standard Model expectation as a function of \MH\ 
is shown.}
\label{fig_s2tw}
}
\hfill
\centering\parbox[t]{.48\textwidth}{\centering\leavevmode
\epsfysize=6cm
\centerline{\epsffile{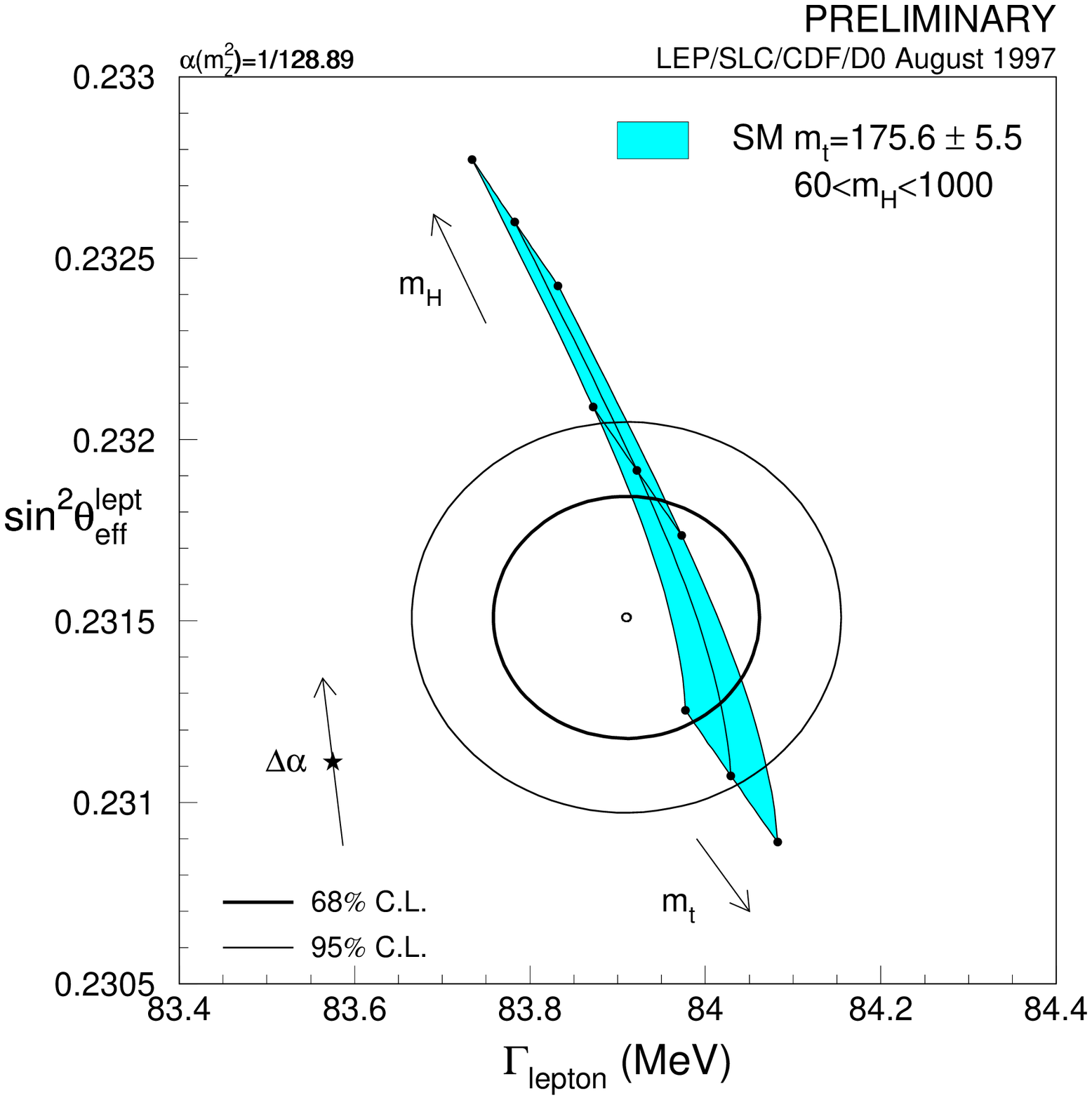}}
\caption{Combined measurements of \stlef\ and $\Gamma_{\ell}$, 
compared with the Standard Model expectation with and without electroweak
radiative corrections.}
\label{fig_s2twrl}
}
\end{figure}

The various measurements of polarizations and  asymmetries at LEP and SLD
can be interpreted, in the context of the Standard Model, as 
measurements of the effective electroweak mixing parameter
\stlef.  In fig.~\ref{fig_s2tw} we show a comparison of the various
determinations of \stlef.  The values are not incompatible with a common
value ($\chi^2$/dof=12.6/6), though it should be noted that the
two most precise determinations (from $A_{\mathrm{LR}}$ and the 
forward-backward b-quark asymmetry) show the largest discrepancies
from the mean.  
In fig.~\ref{fig_s2twrl} we show the measured values of 
\stlef\ and the leptonic width $\Gamma_{\ell}$, which are in good agreement 
with the Standard Model expectation.  The star indicates the prediction
if only photonic radiative corrections are applied, with the arrow
showing the non-negligible uncertainty induced by the running of the 
electromagnetic coupling.  The data clearly demonstrate the need for 
electroweak radiative corrections, and their sensitivity to the 
Higgs mass \MH\ is also evident.
           
\section{Fermion pair production at LEP~II}
Fermion-pair production at energies well above the Z resonance
is characterized by a tendency for {\em radiative return} to the Z
through the emission of one or more photons from the initial state.
The main physics interest, and the greatest sensitivity to new physics,
lies in the events with only a small amount of initial state radiation.
{\em Non-radiative} events are selected by imposing a cut on
the effective c.m. energy of the fermion-pair, $\sqrt{s'}$, which can be
reconstructed from the event kinematics.  Measurements now exist 
of fermion pair cross-sections and forward-backward asymmetries
up to 183~GeV (see~\cite{cpw} for a compilation and detailed references).  

The measurements are all in excellent agreement with the Standard Model.
They may be interpreted in various ways -- either to constrain parameters
of the Standard Model, or to place limits on new physics (for more
details, see~\cite{cpw}).  For example, the hadronic cross-section 
at LEP~II can be used to constrain $\PZ-\gamma$ interference.  
The data taken on the Z peak constrain this interference only
weakly, so it is normally fixed to its Standard Model expectation.
A more model independent interpretation of the data can be performed
using the S-matrix formalism~\cite{Smat}, allowing the parameter
$j^{\mathrm{tot}}_{\mathrm{had}}$ which
parametrizes \Pgg-\PZ\ interference in the hadronic cross-section to be free.
By including data above and below the Z, where interference is sizeable,
the precision of the determination of $j^{\mathrm{tot}}_{\mathrm{had}}$, 
and hence of $\MZ$ in this framework, are greatly improved.
The values obtained are
$\MZ=91.1882\pm0.0029$ and $j^{\mathrm{tot}}_{\mathrm{had}}=0.14\pm0.12$ 
(c.f.\ $j^{\mathrm{tot}}_{\mathrm{had}}=0.22$ in the Standard Model).             
\section{Mass of the W boson}
As we shall see below, the precise measurements of the Z boson 
allow the mass of the W boson to be predicted with a precision
of of around $\pm40$~MeV.  A major goal of the LEP~II and Tevatron
programs is to match this precision by direct measurement,
so as to provide a new test of the Standard Model.

\begin{figure}
\centering\parbox[t]{.48\textwidth}{\centering\leavevmode
\epsfysize=6cm
\centerline{\epsffile{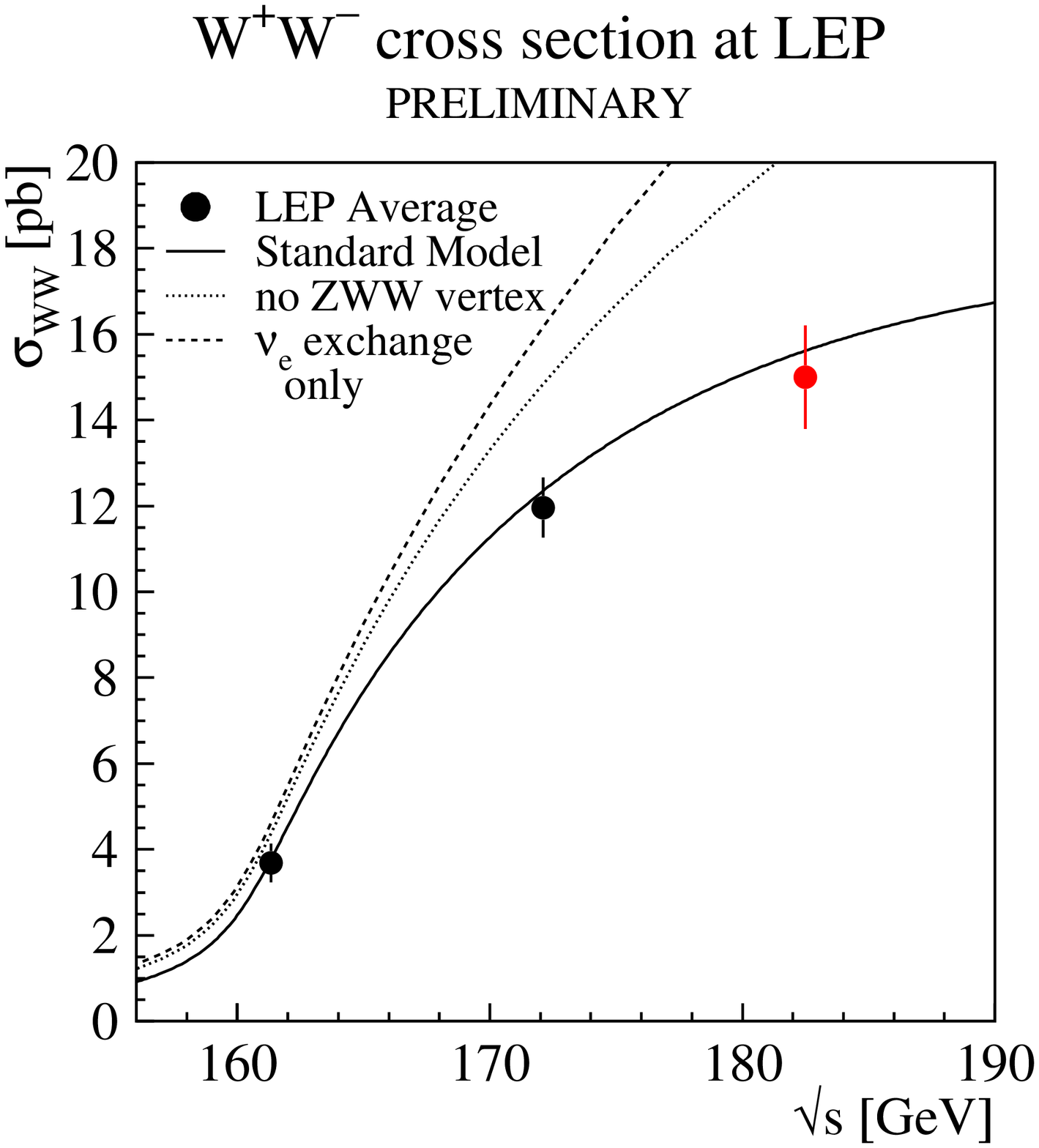}}
\caption{Measurements of the $\Pep\Pem\rightarrow\PWp\PWm$
cross-section at LEP~II.}
\label{fig_wwxs}
}
\hfill
\centering\parbox[t]{.48\textwidth}{\centering\leavevmode
\epsfysize=6cm
\centerline{\epsffile[0 350 600 800]{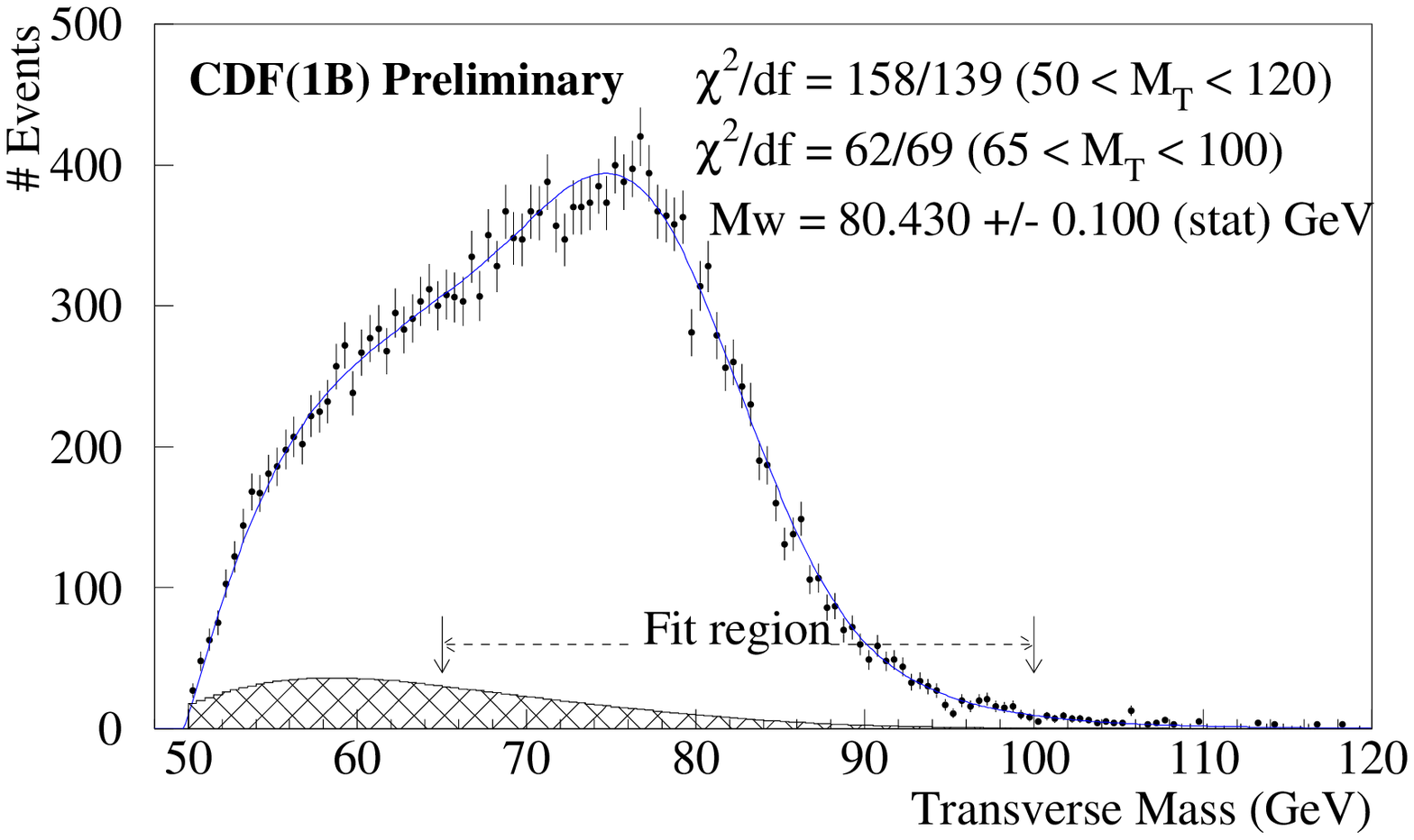}}
\caption{Transverse mass distribution in CDF, from which the 
W mass can be extracted.}
\label{fig_cdfmt}
}
\end{figure}

At LEP~II, \PWp\PWm\ pairs are produced either via $s$-channel \PZ/\Pgg\ 
or $t$-channel neutrino exchange.  The first runs at LEP~II were at 
161~GeV, just above \PWp\PWm\ threshold.  At this energy, the 
\PWp\PWm\ cross-section is very sensitive to the W mass.
The \PWp\PWm\ cross-section, averaged over all four LEP 
experiments~\cite{MW161},
is shown in fig.~\ref{fig_wwxs}.  From the cross-section at 161~GeV,
the W mass is obtained as $\MW=80.40\pm0.22$~GeV, where the error is
predominantly statistical.

At higher energies, the cross-section is much less sensitive to the W mass, 
and the better technique is to reconstruct the W mass directly 
from the invariant mass of its decay products.  
The final states $\PWp\PWm\rightarrow\Pq\Paq\Pq\Paq$ and 
$\PWp\PWm\rightarrow\Pq\Paq\ell\nu_{\ell}$ can both be used for this purpose. 
Kinematic fit techniques, imposing energy and momentum conservation and 
equality of the two W masses in each event, are used to improve the mass 
resolution.  The results obtained after averaging the measurements from
each LEP experiment~\cite{MW172,fasso}, are:
\begin{center}
\begin{tabular}{r@{\hspace{2em}}c}
\hline
Channel         & \MW\ /GeV \\
\hline
$\PWp\PWm\rightarrow\Pq\Paq\Pq\Paq$        & 80.62$\pm$0.26 \\ 
$\PWp\PWm\rightarrow\Pq\Paq\ell\nu_{\ell}$ & 80.46$\pm$0.24 \\
Combined (172~GeV)                         & 80.53$\pm$0.18 \\
\hline
LEP 161 and 172~GeV                        & 80.48$\pm$0.14 \\
\hline
\end{tabular}
\end{center}

In this direct reconstruction approach, the $\Pq\Paq\Pq\Paq$ final state
is potentially more problematical than $\Pq\Paq\ell\nu_{\ell}$ because it
can be affected by hadronic final state interaction effects.   
These can arise because the two Ws typically decay so close together that
they are within the range of the strong interaction.  It has been suggested
that {\em colour reconnection} effects~\cite{yb2} 
could bias the reconstructed W mass
in the $\Pq\Paq\Pq\Paq$ channel by several hundred MeV.  
However, the models which predict the largest effect~\cite{eg} also 
predict other observable effects, such as a $\sim 10\%$ reduction in
the hadron multiplicity in the $\Pq\Paq\Pq\Paq$ case.  
Data already exist~\cite{pr307} comparing the hadronic W decay multiplicity in the 
$\Pq\Paq\Pq\Paq$ and $\Pq\Paq\ell\nu$ final states, yielding a ratio
of 1.04$\pm$0.03.  At first sight this appears inconsistent with the most
extreme colour reconnection model, though caution is needed because
the models used to correct the data do not include the colour reconnection
effect.  Another possible problem could result from Bose-Einstein 
correlations between pions from different Ws.  
Data from LEP so far~\cite{pr717}, 
with large errors, show no evidence for such correlations, consistent with 
the most recent theoretical investigations~\cite{kkm}. 

At the Tevatron, W bosons are produced singly from $\Pq\Paq'$ fusion.
The leptonic W decays $\PW\rightarrow\ell\nu$
($\ell=\Pe/\Pgm$) are used to defeat QCD background, with the 
neutrino inferred from missing momentum.   
The value of \MW\ may be extracted from the distribution of 
transverse mass of $\ell\nu$, as shown in fig.~\ref{fig_cdfmt}.
The current results from CDF and D0 are
80.375$\pm$0.120~GeV and 80.44$\pm$0.11~GeV 
respectively~\cite{MWfnal,cdfw,d0w}.  
The combined W mass measurement from hadron collider experiments
(including UA2) is $80.41\pm0.09$~GeV.

The measurements of \MW\ from hadron colliders and from LEP~II are therefore
in excellent agreement.  The combined ``World Average'' is:
\[  \MW=80.43 \pm 0.08 \;\mathrm{GeV} \]
At present this average is dominated by the Tevatron measurements.
However, if LEP delivers say 50\pbinv\ of data per experiment in 1997, 
the LEP error can be expected to reduce to around 0.08~GeV.
By the end of LEP~II and after the Tevatron upgrade, both LEP and 
Fermilab expect to be able to reach a precision on \MW\ around 0.03--0.04~GeV.

\section{Global Standard Model Fits}

Combined fits of the Standard Model have been performed 
to the measurements of Z decays and the W mass outlined above.
Additional measurements can also be included: the top quark mass
$\Mt=175.6\pm5.5$~GeV~\cite{mt}, the value of 
$1-M^2_{\mathrm{W}}/M^2_{\mathrm{Z}}=0.2254 \pm 0.0037$ from 
$\nu \mathrm{N}$ scattering (which includes a new result from 
CCFR~\cite{ccfr}) and the value of the electromagnetic coupling,
$1/\alpha(\MZ)=128.894\pm0.090$~\cite{ej},
which carries an error because of the need to run it to scale \MZ.
In the fits, $m_{\mathrm{H}}$, $\alphas(\MZ)$ 
and optionally $m_{\mathrm{t}}$ and \MW\ 
are treated as free parameters.  

\begin{figure}
\centering\parbox[t]{.48\textwidth}{\centering\leavevmode
\epsfysize=6cm
\centerline{\epsffile{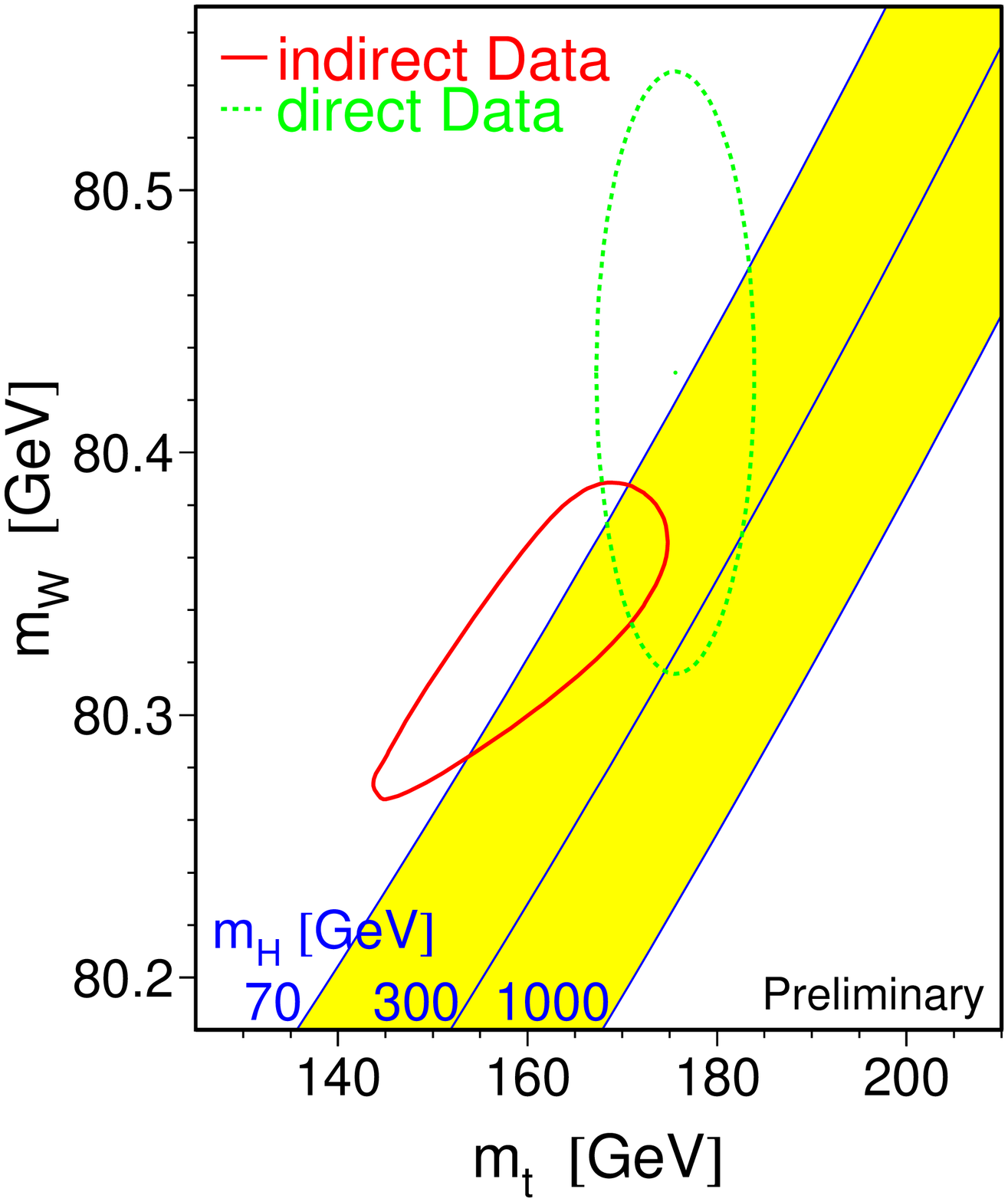}}
\caption{Direct and indirect measurements of \MW\ and \Mt, compared with
the Standard Model for various values of \MH\ (68\% probability contours).}
\label{fig_mwmt}
}
\hfill
\centering\parbox[t]{.48\textwidth}{\centering\leavevmode
\epsfysize=6cm
\centerline{\epsffile{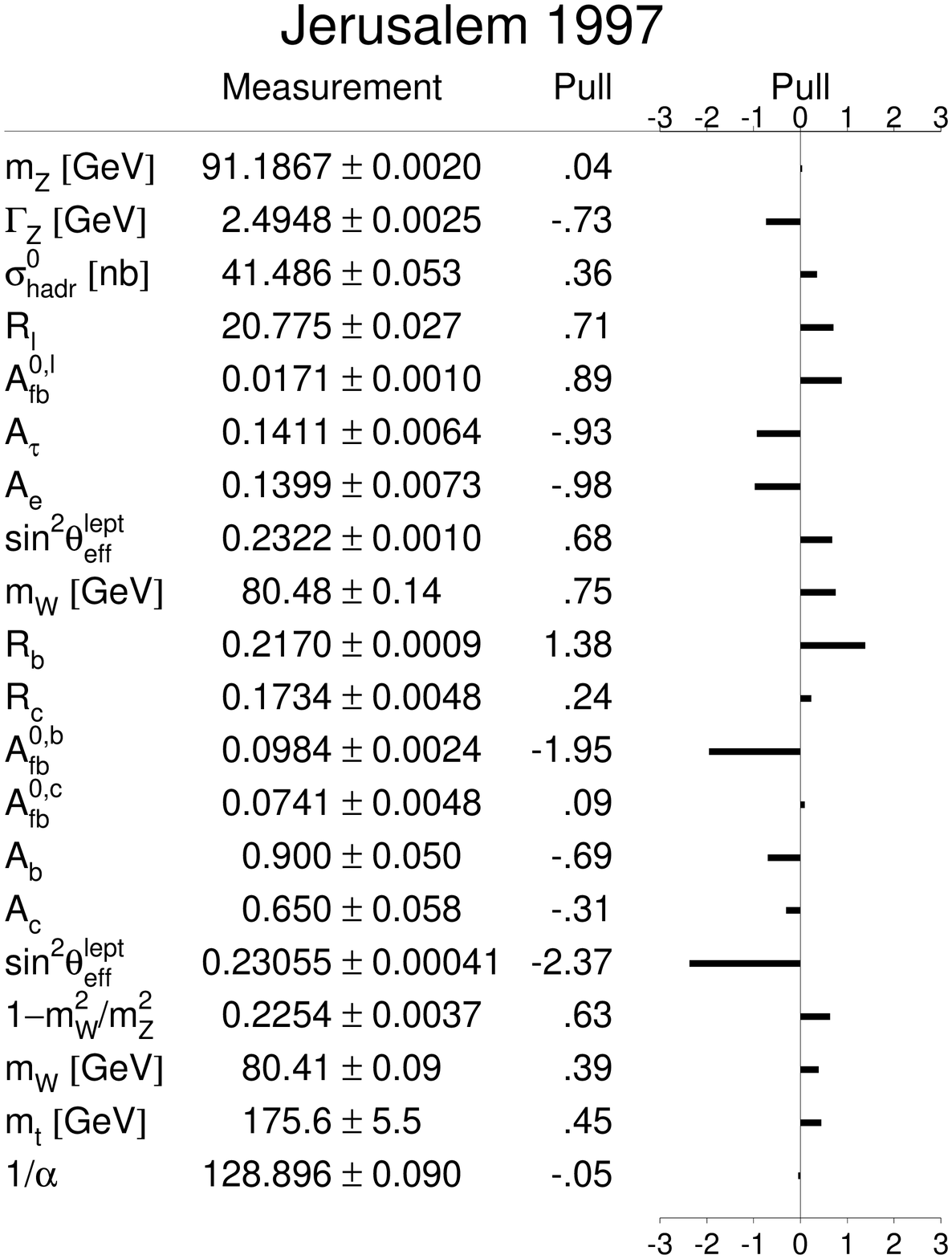}}
\caption{Measured quantities and their pulls in the global
Standard Model fit.}
\label{fig_pulls}
}
\end{figure}

Three fits have been performed, with the results shown below: 
\begin{enumerate}
\item[{\it i)}]
A fit to LEP~I and LEP~II data only.
\item[{\it ii)}]
A fit to all data except the direct measurements of \MW\ and \Mt.
This permits a direct comparison between the direct and indirect 
determinations of these masses, as shown in fig.~\ref{fig_mwmt}.
The two sets of measurements are seen to be consistent.
\item[{\it iii)}]
A fit to all data.  As in the previous two fits, the $\chi^2$/dof value
is excellent, showing that the data are globally compatible with
the Standard Model.  
In fig.~\ref{fig_pulls} we show the input measurements
and the {\em pulls} for this fit, i.e. the difference between 
fitted and measured values divided by the error.  The distribution of pulls
is satisfactory, with only one measurement 
($\sin^2\theta^{\mathrm{lept}}_{\mathrm{eff}}$ from $A_{\mathrm{LR}}$)
more than two standard deviations from the Standard Model.
\end{enumerate}

\begin{center}
\begin{tabular}{l@{\hspace{2em}}c@{\hspace{1em}}c@{\hspace{1em}}c}
\hline
   & {\em i)} LEP (inc. \MW)  & {\em ii)} All but \MW, $m_{\mathrm{t}}$ & 
{\em iii)} All data \\
\hline
$\Mt$ / GeV & $158^{+14}_{-11}$ & $157^{+10}_{-9}$ & $173.1\pm5.4$ \\
$\MH$ / GeV & $83^{+168}_{-49}$ &$41^{+64}_{-21}$  & $115^{+116}_{-66}$ \\
$\log \MH$  &$1.92^{+0.48}_{-0.39}$ &$1.62^{+0.41}_{-0.31}$ &$2.06^{+0.30}_{-0.37}$\\
$\alphas(\MZ)$         &0.121$\pm$0.003 &0.120$\pm$0.003 & 0.120$\pm$0.003\\
$\chi^2$/dof           & 8/9 & 14/12 & 17/15\\
$\MW$ / GeV &80.298$\pm$0.043 &80.329$\pm$0.041 & 80.375$\pm$0.030\\
\hline
\end{tabular}
\end{center}

\begin{figure}

\centering\parbox[t]{.48\textwidth}{\centering\leavevmode
\epsfysize=6cm
\centerline{\epsffile{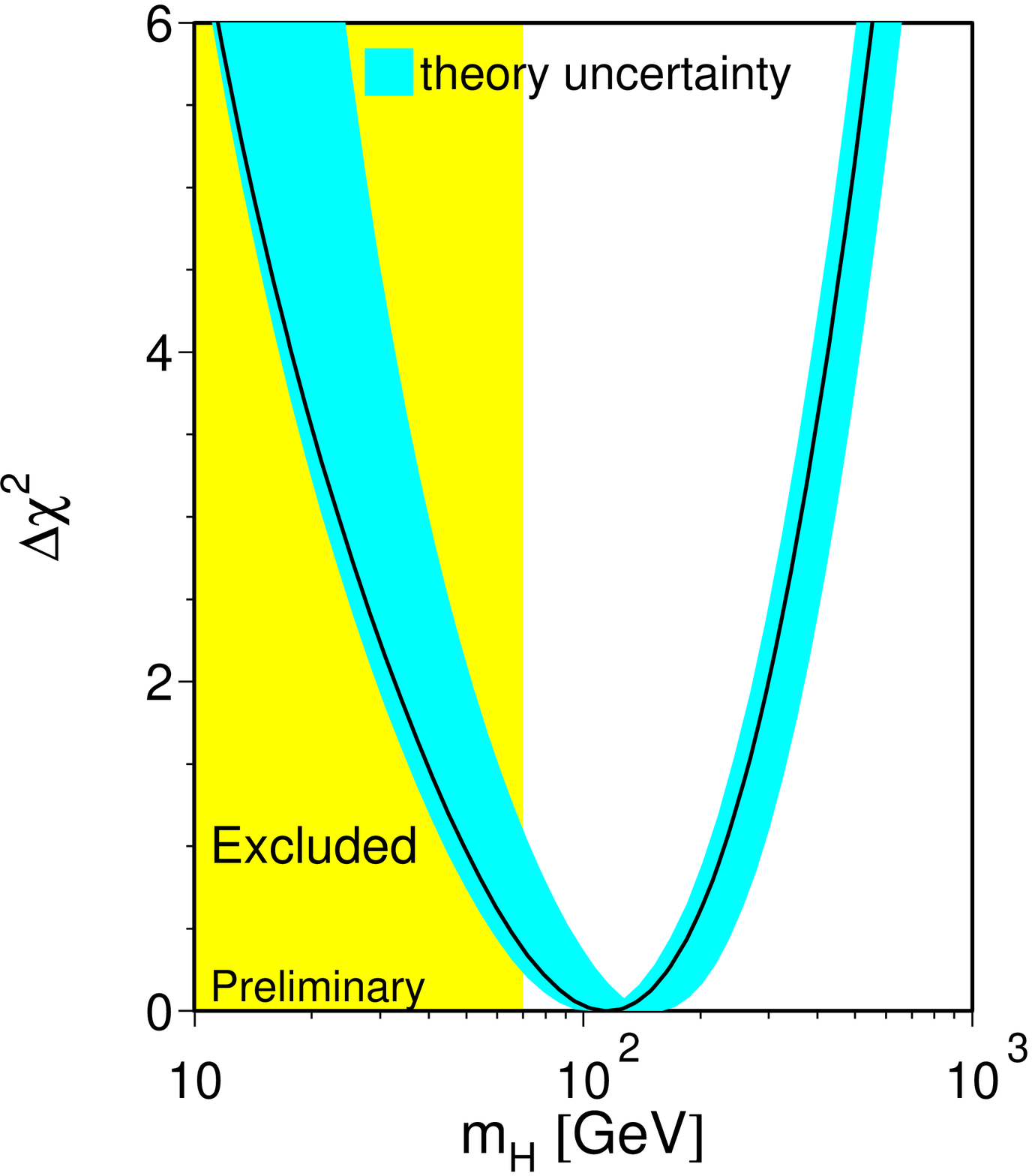}}
\caption{Dependence of $\chi^2$ of the global electroweak fit on
\MH.}
\label{fig_mh}
}
\hfill
\centering\parbox[t]{.48\textwidth}{\centering\leavevmode
\epsfysize=6cm
\centerline{\epsffile{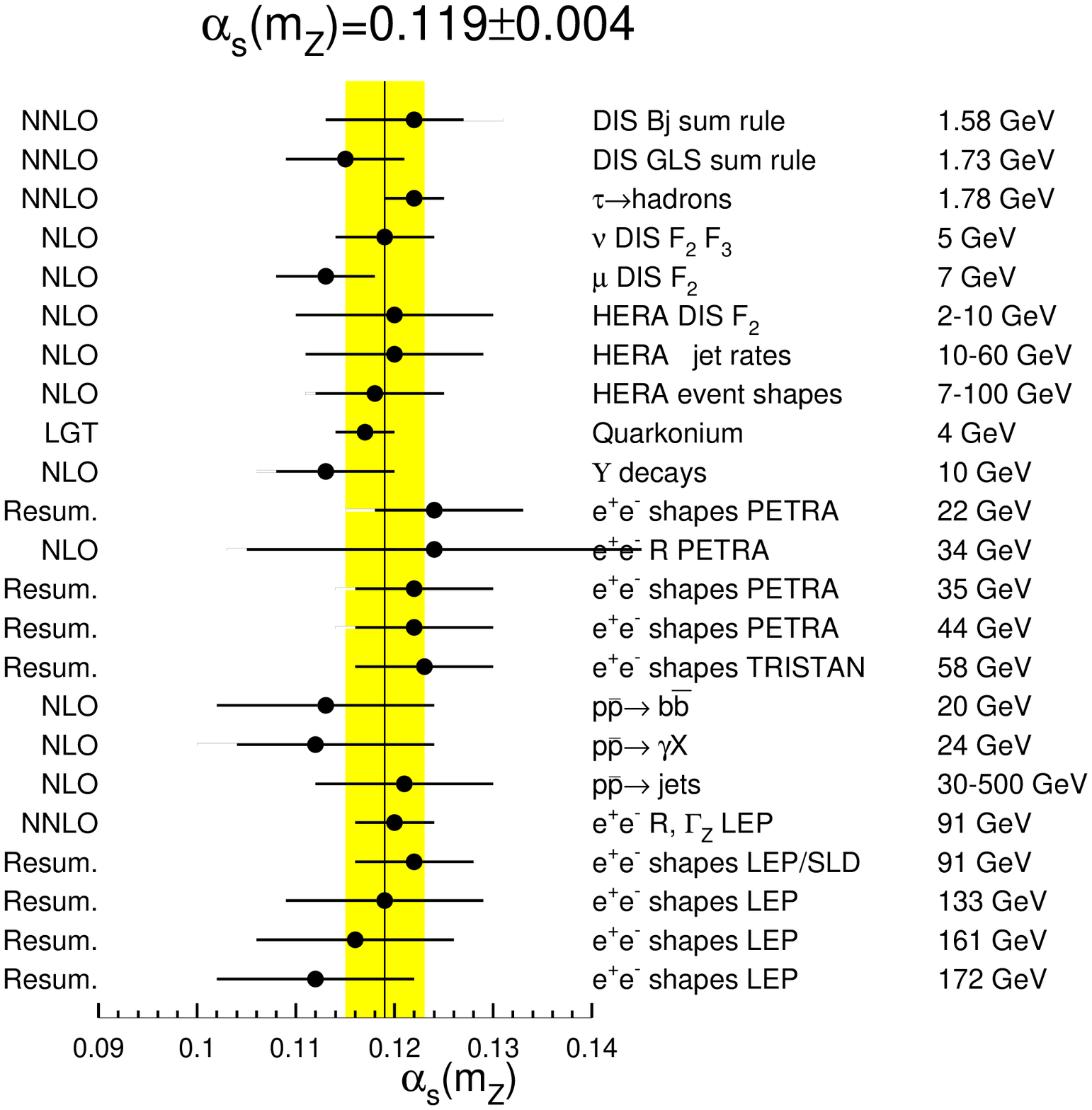}}
\caption{A compilation of recent measurements of \alphas.}
\label{fig_as}
}
\end{figure}

The results of the fits provide an indirect estimate of the mass of the Higgs
boson, \MH.  The most precise estimate comes from the fit to all data,
yielding $\MH=115^{+116}_{-66}$~GeV.  
The two most discrepant measurements in the Standard Model fit tend to
pull \MH\ in opposite directions, $A_{\mathrm{LR}}$ favouring a low \MH, 
and $A_{\mathrm{FB}}^{\mathrm{b}}$ preferring a high value.
The dependence on \MH\ of the 
difference between $\chi^2$ and its minimum value is shown in 
fig.~\ref{fig_mh}.  
The band indicates an estimate of the theoretical
uncertainties resulting from uncomputed higher order terms.  
Taking this into account, an upper limit on \MH\ may be placed:
\[ \MH < 420 \; \mathrm{GeV\;\;  (95\% \; c.l.)} \]
In deriving this limit, the lower mass limit derived from direct searches,
\mbox{$\MH>77$~GeV}~\cite{murray,janot}, has not been taken into account.

The value obtained for $\alphas(\MZ)=0.120\pm0.003$ is one of the most accurate
measurements, even after including a theoretical systematic error of about
$\pm0.002$.  This measurement is compared with other recent 
determinations~\cite{as}
in fig.~\ref{fig_as}.  The measurements display a good level of consistency,
which is a pleasing improvement on the situation a couple of years ago.
A reasonable World Average value is
\[ \alphas(\MZ)=0.119\pm0.004 \]
where the error has been estimated very simply as the r.m.s. deviation of
the measurements from the mean.

\section{Width and branching ratios of the W boson}
\subsection{W Width}
The most precise (but indirect) estimate of the W width is based on the
observed W and Z cross-sections at the Tevatron, using the relation
\[ \frac{\sigma_{\PW} \cdot \mathrm{BR}(\PW\rightarrow\ell\nu)} 
        {\sigma_{\PZ} \cdot \mathrm{BR}(\PZ\rightarrow\ell\ell)} =
   \frac{\sigma_{\PW} \cdot \Gamma(\PW\rightarrow\ell\nu)\cdot\Gamma_{\PZ}} 
        {\sigma_{\PZ} \cdot \Gamma(\PZ\rightarrow\ell\ell)\cdot\Gamma_{\PW}} \]
Taking the cross-sections from QCD, the Z data from LEP, and 
$\mathrm{BR}(\PW\rightarrow\ell\nu)$ from the Standard Model, the 
combined CDF/D0 data yield $\Gamma_{\PW}=2.06\pm0.06$~GeV, 
to be compared with the Standard Model expectation of $2.077\pm0.014$~GeV.  
A more direct measurement can be made from a detailed study of the tail of the 
transverse mass distribution (fig.~\ref{fig_cdfmt}).  The latest result
from CDF~\cite{cdfw} is $\Gamma_{\PW}=2.11\pm0.17\pm0.09$~GeV.
First results from LEP~\cite{MW172}, 
based on direct observation of the W lineshape,
have started to appear:            
\begin{center}
$\Gamma_{\PW}=1.74^{+0.88}_{-0.78}$(stat.)$\pm0.25$(syst.) (L3) \\
$\Gamma_{\PW}=1.30^{+0.62}_{-0.55}$(stat.)$\pm0.18$(syst.) (OPAL)
\end{center} 
At present the errors are not competitive,
but an interesting measurement
should be possible by the end of the 1997 run.
\subsection{W branching ratios and \Vcs}
The observation of \PWp\PWm\ production at LEP permits a direct
determination of the W branching ratios.  The combined results from LEP
are~\cite{burgard}:
\begin{center}
\begin{tabular}{r@{\hspace{2em}}c}
\hline
W decay channel & Branching Ratio (\%) \\
\hline
e$\nu$            & $10.8\pm1.3$    \\
$\mu\nu$          & $ 9.2\pm1.1$    \\
$\tau\nu$         & $12.7\pm1.7$    \\
$\ell\nu$         & $10.9\pm0.6$    \\
Hadrons           & $67.2\pm1.7$    \\
\hline
\end{tabular}
\end{center}

The leptonic results are consistent with lepton universality, though not yet
competitive with results from other processes, such as $\tau$ decays.
The hadronic branching ratio can be related to elements of the CKM matrix:
\[ \frac{B_h}{1-B_h}= \sum_{i=\mathrm{u,c};\;j=\mathrm{d,s,b}} |V_{ij}|^2 
\left(1+\frac{\alphas}{\pi}\right)\]
Amongst these CKM elements, \Vcs\ is by far the least well measured
(\Vcs=1.01$\pm$0.18 from D meson decays).
One can therefore take the other elements from the PDG world averages, 
and infer a value $|\Vcs|=0.96\pm0.08$.

Direct measurements of $\PW\rightarrow\mathrm{c}$ using charm tagging
are also appearing~\cite{pr535}, which yield a more direct determination of 
\Vcs.  The values to date are:
\begin{center}
\mbox{$\Vcs=1.13\pm0.43$(stat.)$\pm0.03$(syst.)} (ALEPH) \\
\mbox{$\Vcs=0.87^{+0.26}_{-0.22}$(stat.)$\pm0.11$(syst.)} (DELPHI)
\end{center}
Although \Vcs\ can be constrained 
much more strongly by the unitarity of the CKM
matrix, these direct measurements will ultimately provide an interesting check.
      
\section{Triple Gauge Couplings}
The WWZ and WW\Pgg\ couplings are predicted by the Standard Model, and 
can be tested by the LEP and Tevatron experiments.  The effective Lagrangian
used to parametrize any anomalous couplings involves
$2\times7$ parameters to describe most general Lorentz invariant WWV 
(V=Z,\Pgg) vertices.  By assuming C, P, and electromagnetic gauge 
invariance, this can be reduced to a more practicable set of five parameters:
$\lambda_{\gamma}$, $\lambda_{\PZ}$ (=0 in Standard Model) and
$\kappa_{\gamma}$, $\kappa_{\PZ}$, $g_1^{\PZ}$ (=1 in Standard Model);
$g_1^{\gamma}=1$  results from electromagnetic gauge invariance.
These parameters may be related to the static moments of the \PW:
\begin{center}
\begin{tabular}{rr@{=}l}
 Charge\hspace{1em} & $Q_W$ & $e g_1^{\gamma}$ \\
 Magnetic dipole moment\hspace{1em} & $\mu_W$ & 
$(e/2m_W)(g_1^{\gamma}+\kappa_{\gamma}+\lambda_{\gamma})$ \\
Electric Quadrupole moment\hspace{1em} & $q_W$ & 
$-(e/m_W^2)(\kappa_{\gamma}-\lambda_{\gamma})$ 
\end{tabular}
\end{center}

At LEP~II, anomalous values for these couplings generally increase the 
\PWp\PWm\ cross-section.  We see from fig.~\ref{fig_wwxs} that the measured
cross-sections clearly require the existence of both WWZ and WW\Pgg\ couplings.
Anomalous couplings also influence the production angle of the \PWm\
and affect the helicity states, and hence the decay angles, of the $\PW^{\pm}$.
The $\PWp\PWm\rightarrow\Pq\Paq\ell\nu$ final states are particularly 
sensitive, because the lepton charge allows an unambiguous 
assignment of the W charges.  Futher information can be obtained  
from ``single W production'' (i.e.\ $\Pq\Paq\Pe\nu$ final states with only a
single on-shell W) and $\nu\overline{\nu}\gamma$ final states, which 
are particularly sensitive to the WW\Pgg\ vertex.  

\begin{figure}
\centering\parbox[t]{.48\textwidth}{\centering\leavevmode
\epsfysize=6cm
\centerline{\epsffile{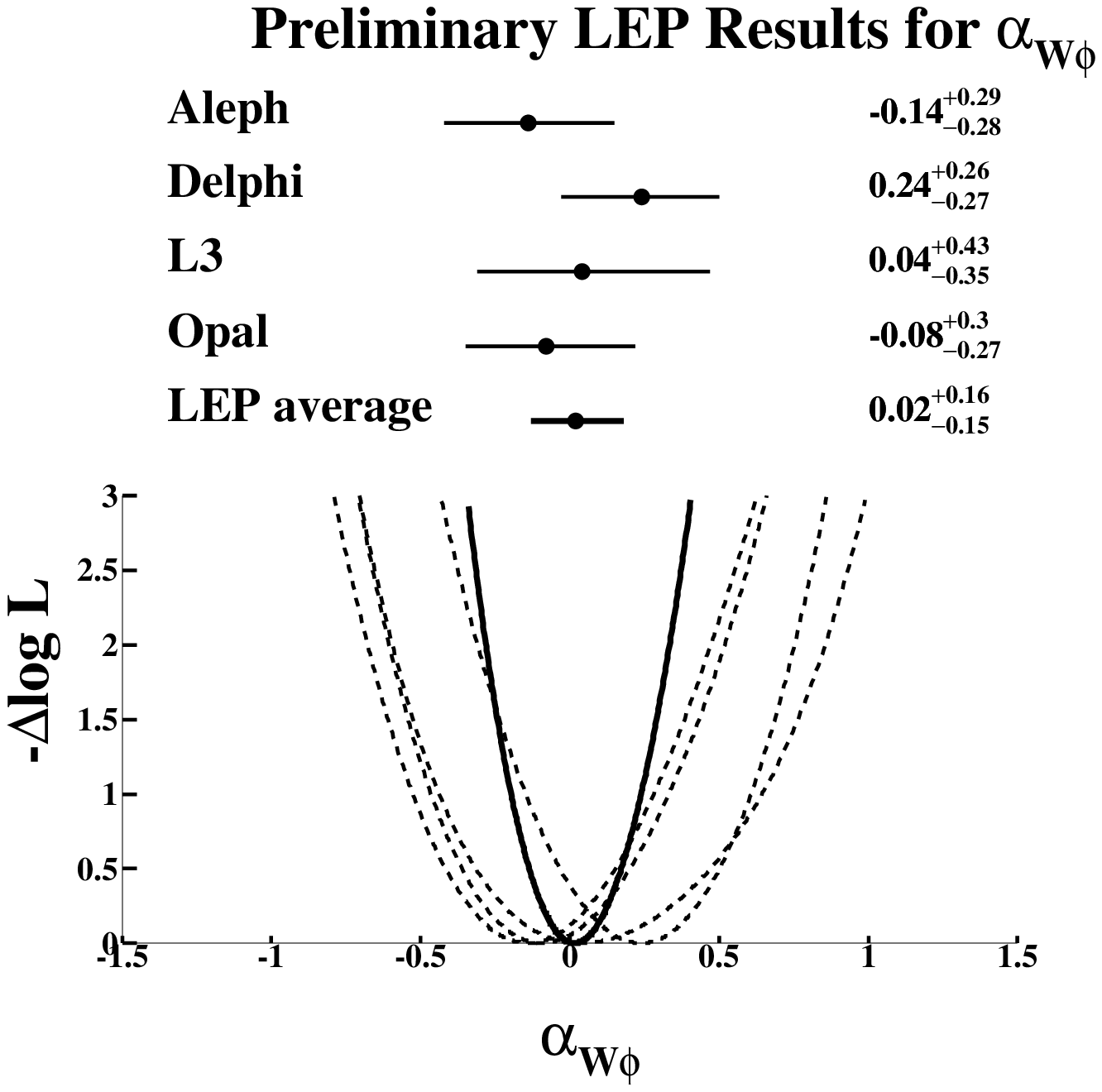}} 
\caption{Likelihood curves for the combined LEP measurement
of $\alpha_{\mathrm{W}\phi}$.}
\label{fig_awf}
}
\hfill
\centering\parbox[t]{.48\textwidth}{\centering\leavevmode
\epsfysize=6cm
\epsffile[0 0 450 450]{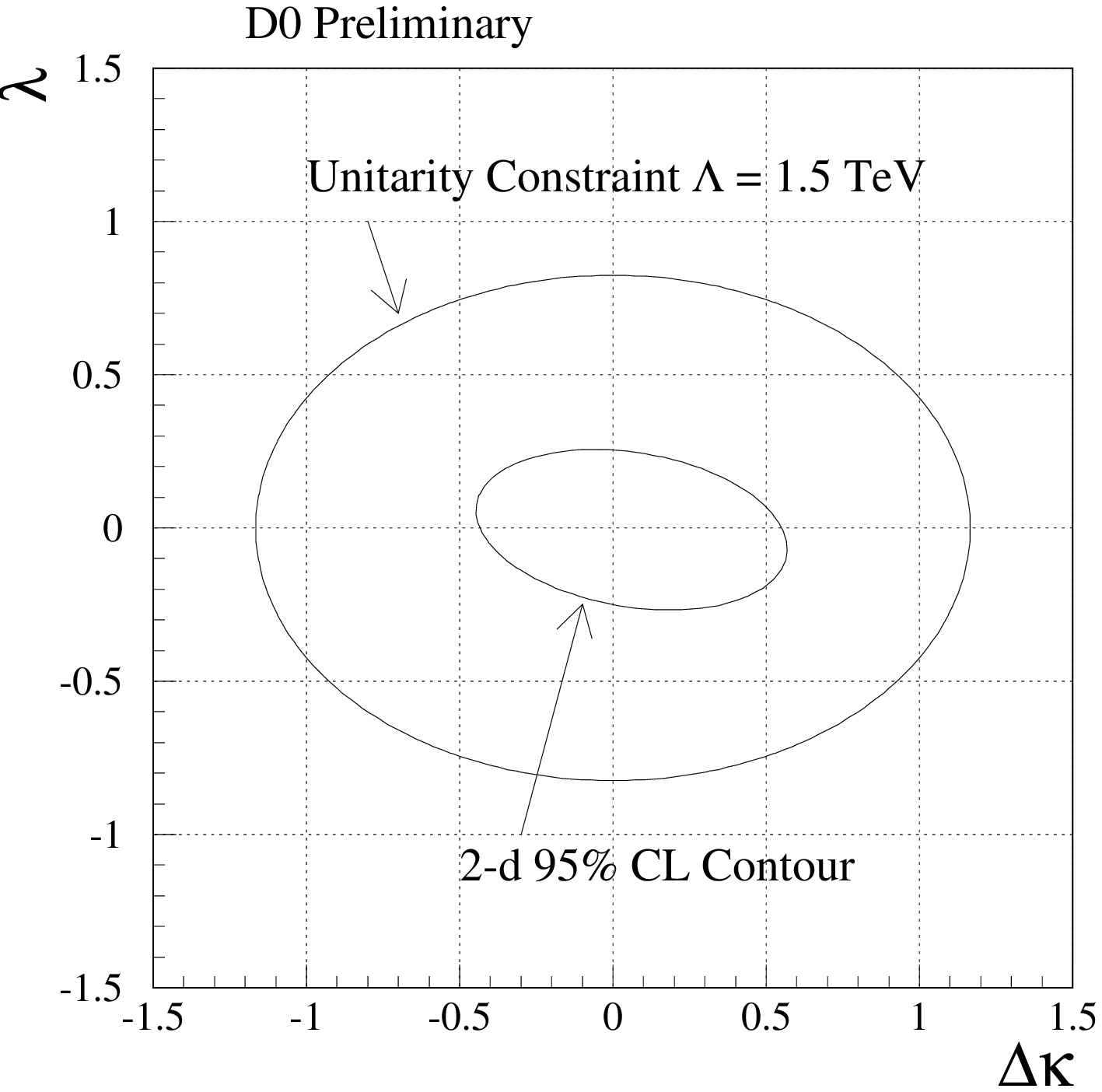} 
\caption{Limits on the couplings $\lambda$ and $\Delta\kappa$ from 
the combined analysis of D0.}
\label{fig_tgcd0}
}
\end{figure}

The precise measurements of the Z already constrain possible anomalous
couplings.  For this reason, the LEP experiments have focussed on 
the following combinations of anomalous couplings, which do
not affect gauge boson propagators at tree level, and are therefore not
already indirectly constrained:
\begin{eqnarray*}
\Delta\kappa_{\Pgg}-\Delta g_1^{\PZ} \cos^2\theta_W & = & \alpha_{B\phi} \\
\Delta g_1^{\PZ} \cos^2\theta_W & = & \alpha_{W\phi}  \\
\lambda_{\PZ}=\lambda_{\Pgg} & = & \alpha_{W} 
\end{eqnarray*}
with the constraint  
$\Delta\kappa_{\PZ}=\Delta g_1^{\PZ}-\Delta\kappa_{\Pgg}\tan^2\theta_W$.
All these couplings should be zero according to the Standard Model.
Combined results from the four experiments have been obtained
by adding likelihood curves from each experiment, taking both 
cross-section and angular information into account, as illustrated in
fig.~\ref{fig_awf}. 
The results are~\cite{weber,leptgc}:
\begin{center}
\begin{tabular}{c@{=}c@{\hspace{2em}}c}
\hline
 \multicolumn{2}{c}{ } & 95\% c.l. limits \\ \hline
$\alpha_{W\phi}$ & $0.02\pm^{0.16}_{0.15}$ &  $-0.28<\alpha<0.33$\\
$\alpha_{W}$     & $0.15\pm^{0.27}_{0.27}$ &  $-0.37<\alpha<0.68$\\
$\alpha_{B\phi}$ & $0.45\pm^{0.56}_{0.67}$ &  $-0.81<\alpha<1.50$\\
\hline
\end{tabular}
\end{center}
Thus no discrepancy from the Standard Model is observed.

In \Pap\Pp\ experiments, information may be gleaned in two ways.
Observation of the rate of $\PW\Pgg$ production, especially for 
high $p_T$ photons, is sensitive to the WW\Pgg\ coupling (and thus
complementary to LEP~II, which is sensitive to WWZ as well).
The results are:
\begin{center}
\begin{tabular}{c@{\hspace{2em}}c@{\hspace{2em}}c}
                 & $\lambda_{\gamma}=0$ & $\Delta\kappa_{\gamma}=0$ \\
\hline
D0~\cite{d0w,pr106}  
& $-0.93<\Delta\kappa_{\gamma}<0.94$ &  $-0.31<\lambda_{\gamma}<0.29$\\
CDF~\cite{cdfw} 
& $-1.8<\Delta\kappa_{\gamma}<2.0$ &  $-0.7<\lambda_{\gamma}<0.6$\\
\hline
\end{tabular}
\end{center}
In addition a few instances of WW or WZ pair production are observed in 
$\ell\nu \ell \nu$ and  $\ell\nu j j$ final states.  
The must stringent limits are obtained by D0~\cite{d0w,pr106} in a combined fit to 
all channels, assuming equal WWZ and WW\Pgg\ couplings:
$-0.33<\Delta\kappa<0.45$  and  $-0.20<\lambda<0.20$, as shown in 
fig.~\ref{fig_tgcd0}.   Where comparison is possible, 
at present the Tevatron limits are typically better
than the LEP~II limits by a factor 2.  This situation should start to change
by the end of the 1997 LEP run.
\section{Summary}
In summary, the electroweak sector of the Standard Model continues 
to stand up to all tests.  The precise electroweak measurements in 
Z decays at LEP~I are coming to an end, and only modest improvements can be expected.  
The distinctive contribution from polarized beam measurements at SLC is 
set to continue.  The results on W physics from LEP~II and the Tevatron 
are starting to appear, and over the next few years a factor $\sim20$ more 
data is anticipated at both machines, to pursue tests of the Standard Model.
\section{Acknowledgements}
Thanks are due to the members of the LEP/SLD Electroweak Working Group,
who have done much of the work of professionally combining the data.
I should also like to thank the many members of ALEPH, DELPHI, L3, OPAL, 
SLD, D0 and CDF who have given me help and information, and the 
organisers of this meeting in Jerusalem for inviting me to speak at such 
an interesting and well run conference.

% ---- Bibliography ----
%

\end{document}